\documentclass[aps,prx,preprint,groupedaddress,longbibliography]{revtex4-2}
\usepackage[T1]{fontenc} 
\usepackage[utf8]{inputenc}                 
\usepackage{microtype}                      
\usepackage{lmodern}

\usepackage{amssymb}                
\usepackage{amsmath}                
\usepackage{amsthm}                 
\usepackage{mathrsfs}               
\usepackage{mathtools}              
\usepackage{comment}
\usepackage{tensor}

\newcommand{\vect}[1]{\boldsymbol{\mathbf{#1}}}     

\let\temp\epsilon                                
\let\epsilon\epsilon
\let\epsilon\temp

\usepackage[bitstream-charter]{mathdesign}
\DeclareSymbolFont{usualmathcal}{OMS}{cmsy}{m}{n}
\DeclareSymbolFontAlphabet{\mathcal}{usualmathcal}

\begin{document}
	\title{Restoring time-reversal covariance in relaxed hydrodynamics}
	\author{Andrea Amoretti$^{\, 1,2}$, Daniel K. Brattan$^{\, 1,3}$, Luca Martinoia$^{\, 1,2}$ and Ioannis Matthaiakakis$^{\, 1,2}$.}
	\affiliation{\textbf{1} Dipartimento di Fisica, Universit\`a di Genova,
		via Dodecaneso 33, I-16146, Genova, Italy,
	}
	\affiliation{\textbf{2} I.N.F.N. - Sezione di Genova, via Dodecaneso 33, I-16146, Genova, Italy,}
	\affiliation{\textbf{3} CPHT, CNRS, École polytechnique, Institut Polytechnique de Paris, 91120 Palaiseau, France.
}	
	\email{andrea.amoretti@ge.infn.it}
	\email{danny.brattan@gmail.com}
	\email{luca.martinoia@ge.infn.it}
	\email{\\ioannis.matthaiakakis@edu.unige.it}
\begin{abstract}
		{ In hydrodynamics, for generic relaxations, the stress tensor and $U(1)$ charge current two-point functions are not time-reversal covariant. This remains true even if the Martin-Kadanoff procedure happens to yield Onsager reciprocal correlators. We consider linearised relativistic hydrodynamics on Minkowski space in the presence of energy, $U(1)$ charge and momentum relaxation. We then show how one can find the minimal relaxed hydrodynamic framework that does yield two-point functions consistent with time-reversal covariance. We claim the same approach naturally applies to boost agnostic hydrodynamics and its limits (e.g. Carrollian, Galilean and Lifshitz fluids).}
\end{abstract}
	
\maketitle
\newpage

\section{Introduction}

{\noindent When we couple a charged fluid to an external electric field in order to compute the fluid's optical conductivities, we find that these conductivities diverge in the strict DC limit $\omega\rightarrow0$ \cite{Hartnoll:theory}. The origin of this divergence traces back to the exact translation symmetry of the theory, which implies that a constant electric field (or temperature gradient) adds momentum to the fluid without bound. The standard approach to solve this problem is the introduction of a small relaxation parameter $\tau_p^{-1}$ which breaks translation invariance and relaxes the total momentum. While momentum is no longer conserved, it can still be a relevant hydrodynamic charge if $\tau_p$ is large enough \cite{Hartnoll:theory} (for some holographic examples see \cite{Andrade:simple_holographic_model,Amoretti:2017xto,Amoretti:2014zha,Amoretti:magneto-transport_holography}).}

{\ In the presence of a $U(1)$ axial anomaly, momentum relaxation is not enough to obtain finite DC conductivities. In this case one must also relax energy and $U(1)$ charge conservation \cite{Landsteiner:negative_magnetoresistivity,Abbasi:magneto-transport,Lucas:hydrodynamic_theory,Landsteiner:anomalous_magnetoconductivity_holography}. The inclusion of these relaxation terms is natural, in the sense that energy relaxation is a fact of certain condensed matter processes and charge relaxation may appear when the corresponding symmetry is only approximate, see e.g. \cite{Pongsangangan:hydrodynamics_charged,Fritz:hydrodynamic_electronic_transport,Narozhny:hydrodynamic_approach,Gall:electronic_viscosity} for a kinetic theory derivation and \cite{Amoretti:non_dissipative} for a discussion of relaxed non-dissipative hydrodynamics.}
	
{\ At the computational level, there are two main approaches to obtaining the response functions and their corresponding conductivities in linearised hydrodynamics: the variational or background field approach \cite{Kovtun:lectures} and the  the Martin-Kadanoff or canonical approach \cite{Kadanoff:hydrodynamics}. In the first the fluid is perturbed by spacetime and gauge field fluctuations that couple to the full stress-energy tensor and current respectively, while in the second sources only for the conserved charge densities are introduced.}
	
{\ Because the relaxation terms explicitly break the Lorentz symmetry of the system, the variational approach is usually deemed impractical for computing the Green's functions\footnote{Notice that it is not possible to simply swap each $\omega$ with $\omega+\frac{i}{\tau}$ in the Green's functions.} because it is impossible to write the relaxation terms covariantly. On the other hand, the Martin-Kadanoff method works well in the presence of relaxation terms, but comes short on other aspects. In particular: (i) it does not give access to all the Green functions, but only to those related to the thermodynamic charges\footnote{This is even more relevant in the presence of energy and charge relaxation, because in this scenario we cannot naively use Ward identities to relate e.g. $\langle J^t J^t\rangle$ to $\langle J^i J^i\rangle$.} and (ii) in more general cases, e.g. when there are strong electric and magnetic fields that polarize the fluid, it might not be straightforward to obtain the relations between the charges and their thermodynamic sources. For these reasons, it is of interest to find a unique prescription to study the response functions of fluids  with weakly relaxed charges via the variational method.}
	
{\ In this paper we present a prescription that allows one to compute the two-point functions of relaxed hydrodynamics via the background field method. Namely, we consider a particular example, relaxed relativistic hydrodynamics with a $U(1)$ charge, and introduce all the possible source terms (metric and gauge field fluctuations) to the conservation equations at order one in fluctuations and order zero in derivatives. We find that by simply requiring time-reversal covariance of the microscopic theory we can completely constrain all the extra parameters we introduce. The relaxation terms that survive this procedure exactly reproduce the Green's functions obtained via the Martin-Kadanoff approach up to contact terms.}

{\ The paper is structured as follows: in section \ref{sec:Martin-Kadanoff} we discuss the general properties of constant relaxations in the Martin-Kadanoff formalism for an example $U(1)$ charged relativistic fluid and obtain a set of very general constraints based on Onsager reciprocity, positivity of entropy production and linearised stability. In section \ref{sec:variational_approach} we implement the same relaxations in the variational approach and, by imposing time-reversal covariance on the correlators, show how to obtain consistent results between the prescriptions.}

\section{The Martin-Kadanoff approach and Onsager reciprocity}\label{sec:Martin-Kadanoff}

{\noindent In the present section, we consider the Martin-Kadanoff procedure in the presence of the most general sources of relaxation linear in the fluctuation fields. This approach gives a subset of the possible two-point functions of the stress tensor and $U(1)$ charge current. Subsequently, we impose Onsager reciprocity on the obtained Green's functions alongside the second law of thermodynamics. This naturally leads to strong constraints on the allowed relaxation terms. Having developed this framework, in the next section we shall explore how to make this effective description time-reversal covariant for all the two-point functions of the currents one can calculate in hydrodynamics.}

{\ Our starting point is the linearised hydrodynamic equations of motion, Fourier transformed in the spatial directions. They can generically be brought into the following form
\begin{equation}
	\label{Eq:MKequation}
	\partial_t\varphi_a(t,\vect{k})+M_{ab}(\vect{k})\varphi_b(t,\vect{k})=0~.
\end{equation}
By $\varphi_a$ we denote the fluctuations of the conserved charges, while $M_{ab}$ is the hydrodynamic matrix, whose specific expression depends on the constitutive relations and the equations of motion. Given $M_{ab}$ we can write down an explicit formula for the retarded Green's function $G^R_{ab}$, namely \cite{Kovtun:lectures}
\begin{equation}
\label{Eq:RetardedG}
G^R_{ab}(z, \mathbf{k}) = -(1 + izK^{-1}(z,\mathbf{k}))_{ac}\chi_{cb}~,
\end{equation}
where we defined $K_{ab} = -iz\delta_{ab} + M_{ab}(\mathbf{k})$ and $\chi_{ab}$ is the thermodynamic susceptibility matrix
\begin{equation}
	\chi_{ab}=\frac{\partial\varphi_a}{\partial\lambda_b}~.
\end{equation}
We denote by $\lambda_a$ the sources conjugate to the charges.}

{\ Following \cite{Kovtun:lectures}, we impose time-reversal covariance of the microscopic theory by requiring
\begin{equation}\label{eqn:Onsager_relations}
	G_{ab}^R(\omega,\vect{k})=\eta_a\eta_bG_{ba}^R(\omega,-\vect{k})~,
\end{equation}
where $\eta_a$ is the time-reversal eigenvalue for the field $\varphi_a$. Enforcing time-reversal using \eqref{Eq:RetardedG}, leads to the constraint\footnote{Recall the susceptibility matrix is symmetric, $\chi_{ab} = \chi_{ba}$.}
\begin{equation}\label{eqn:Onsagerhydrodynamics}
	\chi SM^T(-\vect{k})=M(\vect{k})\chi S~,
\end{equation}
with $S=\text{diag}(\eta_1,\eta_2,\dots)$ the matrix of time-reversal eigenvalues of the $\varphi$s. In writing \eqref{eqn:Onsagerhydrodynamics}, we have assumed that there are no explicit parameters $B$, such as the magnetic field, that break time-reversal invariance of the microscopic theory. In such cases one may be able to extend \eqref{eqn:Onsagerhydrodynamics} to a relationship between different theories where the parameter $B$ also transforms appropriately under time reversal. For example, in the case of a constant magnetic field we take $B \rightarrow -B$ under time reversal.}

{\ The above applies for general hydrodynamic theories; now we specialise our discussion to relativistic hydrodynamics. We consider a charged relativistic fluid at temperature $T$ and chemical potential $\mu$ propagating in Minkowski spacetime. Its energy-momentum tensor $T^{\mu\nu}$ and electric current $J^\mu$ are given in the Landau frame by \cite{Kovtun:lectures}
\begin{subequations}
	\label{Eq:NonlinearhydroMink}
	\begin{align}
		T^{\mu \nu} &= \epsilon u^\mu u^\nu+P \Delta^{\mu\nu}-\eta\Delta^{\mu\alpha}\Delta^{\nu\beta}\left(\partial_\alpha u_\beta+\partial_\beta u_\alpha-\frac{2}{d}\eta_{\alpha\beta}\partial_\lambda u^\lambda\right)-\zeta\partial_\mu u^\mu+\mathcal{O}(\partial^2) \; , \\
		J^\mu &= nu^\mu-\sigma\Delta^{\mu\nu}T\partial_\nu\frac{\mu}{T}+\mathcal{O}(\partial^2) \; , 
	\end{align}
\end{subequations}
with $\Delta^{\mu\nu} = \eta^{\mu\nu} + u^\mu u^\nu$, the projector normal to the velocity profile $u^\mu$, $\eta^{\mu\nu} = {\rm diag}(-1,1,1...1)$ and $d$ the number of spatial dimensions.} 

{\ The hydrodynamic equations corresponding to $T^{\mu\nu}$, $J^\mu$ are the energy-momentum and charge conservation equations, $\partial_\mu T^{\mu\nu} = 0 = \partial_\mu J^\mu$. We linearize these equations around a state with constant energy density, number density and zero spatial velocity
\begin{align}
\label{Eq:BackgroundDef}
&u^\mu = u^\mu_{(0)} + \delta u^\mu~, \qquad \epsilon = \epsilon_{(0)} + \delta \epsilon~, \qquad n= n_{(0)} + \delta n~
\end{align}
with $u^{\mu}_{(0)}=(-1,0,\ldots,0)$ in a Cartesian coordinate system. Subsequently we add relaxation terms\footnote{Our relaxations will break Lorentz invariance and one may in principle expect new transport coefficients to appear depending on how the microscopic theory couples to processes responsible for breaking this symmetry. These new coefficients will not change results related to our relaxations and we can, without loss of generality, assume that they happen to be zero for our fluid henceforth.}  that are linear in the charge fluctuations to find
	\begin{subequations}\label{eqn:equations_of_motion}
		\begin{align}
		\partial_{t} \delta \epsilon + \left( \epsilon_{(0)} + P_{(0)} \right) \partial_{i} \delta v^{i} &= - \left( \frac{1}{\tau_{\epsilon \epsilon}} \delta \epsilon +  \frac{1}{\tau_{\epsilon n}} \delta n \right) \; , \\
		\partial_{t} \delta n + n_{(0)}\partial_i\delta v^i -\sigma T_{(0)}\partial_i^2\left(\delta\frac{\mu}{T}\right) &= - \left( \frac{1}{\tau_{n \epsilon}} \delta \epsilon +  \frac{1}{\tau_{nn}} \delta n \right) \; , \\
		\partial_{t} \delta p^{i} +\partial^i\delta P-\eta\left(\partial_j^2\delta v^i+\frac{1}{3}\partial^i\partial_j\delta v^j\right)-\zeta\partial_i\partial_j\delta v^j &= - \frac{1}{\tau_{p}} \delta p^{i} \; , 
	\end{align}
\end{subequations}
with $\delta p_i = (\epsilon_{(0)} + P_{(0)}) \delta v_i$. These equations \eqref{eqn:equations_of_motion} correspond to the (non-)conservation of energy, charge and momentum respectively. The energy and charge relaxation terms are defined intuitively in terms of their corresponding relaxation times. In particular, $\tau_p$ is the momentum relaxation time while $\tau_{\epsilon\epsilon}$ and $\tau_{nn}$ are the energy and charge relaxations times. The terms $\tau_{n\epsilon}$, $\tau_{\epsilon n}$ are mixed relaxations, which as written do not have the units of time, and one can find an example of such terms in \cite{Lucas:hydrodynamic_theory}. We could also consider generalizing momentum relaxation to mix different momentum components however, this type of relaxation can always be reduced to the one found in \eqref{eqn:equations_of_motion} (see section 2.4 of \cite{Amoretti:non_dissipative}). Thus, we have added to the RHSs of \eqref{eqn:equations_of_motion} all possible relaxation terms linear in fluctuations and without explicit derivatives. Subsequently, we take the relaxations times to be small ($\tau$ large) by which we mean, in the linearised regime, that $\tau^{-1}\sim\mathcal{O}(\partial)$.}

{\ Before we discuss in detail the constraints on relaxation stemming from time reversal and the second law, we note that a heuristic thermodynamic argument already suggests the existence of such constraints. Following \cite{Landsteiner:negative_magnetoresistivity}, consider a scattering event between the fluid microscopic constituents and some impurities/defects that relax the energy (or charge). Because of the scatterings, the fluid loses $\delta\epsilon$ energy in a time $\tau_{\epsilon\epsilon}$. However, this implies that it also loses charge $\delta n=\frac{\partial n}{\partial\epsilon}\delta\epsilon=\frac{\delta\epsilon}{\mu_{(0)}}$ in the same time interval. An analogous argument tells us that if the fluid loses $\delta n$ charge, it also loses $\delta\epsilon=\mu_{(0)}\delta n$ energy. Hence we see that in general energy and charge relaxations are strongly intertwined and we can already put forward the ansatz 
\begin{equation}\label{eqn:physical_ansatz}
	\frac{\mu_{(0)}}{\tau_{nn}}=\frac{1}{\tau_{\epsilon n}} \; , \qquad\qquad\frac{1}{\tau_{\epsilon\epsilon}}=\frac{\mu_{(0)}}{\tau_{n\epsilon}} \; .
\end{equation}
We confirm this holds generically for relaxed theories that preserve positivity of entropy production shortly.} 

{\ We impose the Onsager relations to hydrodynamics by requiring \eqref{eqn:Onsagerhydrodynamics} on the equations of motion \eqref{eqn:equations_of_motion}. Taking $\varphi_{a}(t,\vect{k}) = \left( \delta \epsilon, \delta n, \delta p_{i}\right)$ and considering, without loss of generality, $\vect{k}=(k_x,0,0)$ we determine the matrix $M_{ab}$ of \eqref{Eq:MKequation} to be
\begin{equation}
	M=
	\begin{pmatrix}
		\frac{1}{\tau_{\epsilon\epsilon}}   &   \frac{1}{\tau_{\epsilon n}} &   ik_x    &   0   &   0\\
		k_{x}^2\sigma\beta_\epsilon+\frac{1}{\tau_{n\epsilon}} &  k_{x}^2\sigma\beta_n+\frac{1}{\tau_{nn}}    &   \frac{ik_x n_{(0)}}{P_{(0)}+\epsilon_{(0)}}   &   0   &   0\\
		ik_x\frac{\partial P}{\partial\epsilon} &   ik_x\frac{\partial P}{\partial n}   &   \frac{k_{x}^2(3\zeta+4\eta)}{3(P_{(0)}+\epsilon_{(0)})}+\frac{1}{\tau_p} &   0   &   0   \\
		0   &   0   &   0   &   \frac{k_{x}^2\eta}{P_{(0)}+\epsilon_{(0)}}+\frac{1}{\tau_m}  &   0   \\
		0   &   0   &   0   &   0   &   \frac{k_{x}^2\eta}{P_{(0)}+\epsilon_{(0)}}+\frac{1}{\tau_m}
	\end{pmatrix}
\end{equation}
where we defined $\beta_\epsilon=\frac{\partial\mu}{\partial\epsilon}-\frac{\mu_{(0)}}{T_{(0)}}\frac{\partial T}{\partial\epsilon}$, $\beta_n=\frac{\partial\mu}{\partial n}-\frac{\mu_{(0)}}{T_{(0)}}\frac{\partial T}{\partial n}$ and the matrix of the time-reversal eigenvalue is $S = \mathrm{diag}(1,1,-1,-1,-1)$. We find
\begin{align}\label{eqn:Onsager_general}
	\frac{\chi_{\epsilon\epsilon}}{\tau_{n\epsilon}}-\frac{\chi_{\epsilon n}}{\tau_{\epsilon\epsilon}}+\frac{\chi_{n\epsilon}}{\tau_{nn}}-\frac{\chi_{nn}}{\tau_{\epsilon n}}=0
\end{align}
where the susceptibilities are 
	\begin{align}
		\chi_{n\epsilon} =\chi_{\epsilon n}=\frac{\partial\epsilon}{\partial\mu}= T_{(0)} \frac{\partial n}{\partial T} + \mu_{(0)} \frac{\partial n}{\partial \mu} \; , \qquad
		\chi_{nn} =\frac{\partial n}{\partial\mu} \; , \qquad
		\chi_{\epsilon\epsilon} =T_{(0)}\frac{\partial\epsilon}{\partial T}+\mu_{(0)}\frac{\partial\epsilon}{\partial\mu} \; .
	\end{align}
The thermodynamic derivatives are taken in the grand canonical ensemble, at fixed $\mu$ or $T$ respectively. We can see then that if we set $\tau_{n\epsilon}^{-1}=\tau_{\epsilon n}^{-1} =0$ leaving only the pure charge and energy relaxations typically encountered in the literature, we must impose $\tau_{\epsilon \epsilon}=\tau_{nn}$. This relation is known \cite{Abbasi:magneto-transport}, but we shall now see that this truncated relaxation is at odds with the second law of thermodynamics.}

{\ The linearised entropy current is given by
	\begin{subequations}
	\begin{eqnarray}
		\delta s^{\mu} &=& \delta s^{\mu}_{\mathrm{can.}} + \delta s^{\mu}_{\mathrm{eq.}} \; , \\
		s^{\mu}_{\mathrm{can.}} &=& \frac{1}{T} \left( P u^{\mu} + T\indices{^{\mu \nu}} u_{\nu} - \mu J^{\mu} \right) \; ,
	\end{eqnarray}
	\end{subequations}
where $s^{\mu}_{\mathrm{eq}.}$ is present to ensure that the hydrodynamic equations vanish upon imposition of hydrostaticity and is zero at the relevant order in our current situation. It is then not difficult to show that the divergence of this current takes the form 
\begin{equation}
	\label{Eq:LowestEntropy}
	T_{(0)}\partial_\mu \delta s^\mu=\delta\epsilon\left(\frac{\mu_{(0)}}{\tau_{n\epsilon}}-\frac{1}{\tau_{\epsilon\epsilon}}\right)+\delta n\left(\frac{\mu_{(0)}}{\tau_{nn}}-\frac{1}{\tau_{\epsilon n}}\right) + \mathcal{O}(\delta^2,\partial^2) \; . 
\end{equation}
Positivity of entropy production requires that the divergence of the entropy current be positive definite on any (including the linearised) solution of the hydrodynamic equations of motion. As each of the fluctuations $\delta\epsilon$ and $\delta n$ in \eqref{Eq:LowestEntropy} are of arbitrary sign, it follows that their coefficients must be zero.\footnote{We could also express the second law in terms of $\delta T$ and $\delta \mu$. This leads to a $2\times 2$ linear system of equations that has a solution with ${\rm det}\chi = 0$. In what follows, we ignore this solution as unphysical.} This gives two constraints on the relaxation rates and hence we obtain \eqref{eqn:physical_ansatz}, confirming that the relaxations of energy and charge are connected for thermodynamic reasons. Subsequently, with these constraints between the relaxation rates due to the second law, we can use  \eqref{eqn:Onsager_general} to find
	\begin{equation}\label{eqn:entropy_constraint_order_one}
		\frac{\partial\epsilon}{\partial T}\frac{1}{\tau_{\epsilon\epsilon}}+\frac{\partial n}{\partial T}\frac{\mu_{(0)}}{\tau_{nn}}=0
	\end{equation}
This leaves a one parameter family of relaxations which we can parameterise by $\tau_{nn}$. Notice that in general, and this will be confirmed by the study of the modes, not all relaxation times must be positive.}

{\ Equations \eqref{Eq:LowestEntropy} and \eqref{eqn:entropy_constraint_order_one} fix all but one of the relaxation terms giving us a one-parameter family that, at least at the linearised level, satisfy positivity of entropy production and Onsager reciprocity. Importantly, we can see that, unless the chemical potential is zero, we will find $\tau_{\epsilon n} \neq 0$ whenever $\tau_{nn} \neq 0$, if we want these properties to hold. Alternatively, one can have more generic relaxation rates if the second law is ignored. In this case, the relaxation rates represent the coupling of a fluid to an open system, rather than some UV degrees of freedom in a more complete quantum theory with a quasi-hydrodynamic limit (see e.g. the discussion in \cite{Amoretti:hydrodynamic_magneto-transport,Amoretti:2021lll,Amoretti:2022acb}).} 

{\ In \eqref{Eq:LowestEntropy} we examined the entropy positivity condition at lowest order in fluctuations. However, when considering linearised hydrodynamics, constraints on transport coefficients can be inferred only by examining the second law at order two in fluctuations. In our case, at order two in fluctuations, the divergence of the entropy current is given by
\begin{align}
	T_{(0)} \partial_\mu \delta s^\mu&=\frac{\delta\epsilon\delta T}{T_{(0)}\tau_{\epsilon\epsilon}}+\frac{\delta n\delta T}{T_{(0)}\tau_{\epsilon n}}+\frac{\delta\mu\delta n}{\tau_{nn}}+\frac{\delta\mu\delta\epsilon}{\tau_{n\epsilon}}-\mu_{(0)}\frac{\delta T\delta n}{T_{(0)}\tau_{nn}}-\mu_{(0)}\frac{\delta T\delta\epsilon}{T_{(0)}\tau_{n\epsilon}}\nonumber\\
	&\quad +\frac{1}{\tau_p}(\epsilon_{(0)} + P_{(0)})\delta v^2 + \sigma \left(\frac{\mu_{(0)}}{T_{(0)}} \partial \delta T- \partial \delta \mu \right)^2  \nonumber \\
	&\quad +\eta\left(\delta\sigma^{ij}\right)^2+\zeta\left(\partial_i\delta v^i\right)^2+ \mathcal{O}(\delta^3,\partial^3)
\end{align}
Positivity of the RHS gives the usual constraints on the transport coefficients $\sigma\geq0, \eta\geq0$ and $\zeta\geq0$. Furthermore we also find $\tau_p\geq0$ as expected for momentum relaxation. We can rewrite the remaining relaxations in terms of fluctuations of just $T$ and $\mu$ to find
\begin{eqnarray}
		\label{Eq:Entropy2}
	    T_{(0)}\partial_\mu \delta s^\mu
	&= & \delta\mu\delta T\left(\frac{\partial\epsilon}{\partial\mu}\frac{1}{T_{(0)}\tau_{\epsilon\epsilon}}+\frac{\partial n}{\partial\mu}\frac{1}{T_{(0)}\tau_{\epsilon n}}+\frac{\partial n}{\partial T}\frac{1}{\tau_{nn}}+\frac{\partial\epsilon}{\partial T}\frac{1}{\tau_{n\epsilon}} -\frac{\mu_{(0)}}{T_{(0)}}\frac{\partial n}{\partial\mu}\frac{1}{\tau_{nn}} \right. \nonumber \\
	&\;& \left. \hphantom{\delta\mu\delta T\left( \right.} -\frac{\mu_{(0)}}{T_{(0)}}\frac{\partial\epsilon}{\partial\mu}\frac{1}{\tau_{n\epsilon}}\right)  +\frac{\delta T^2}{T_{(0)}}\left(\frac{\partial\epsilon}{\partial T}\frac{1}{\tau_{\epsilon\epsilon}}+\frac{\partial n}{\partial T}\frac{1}{\tau_{\epsilon n}}-\frac{\partial n}{\partial T}\frac{\mu_{(0)}}{\tau_{nn}}-\frac{\partial\epsilon}{\partial T}\frac{\mu_{(0)}}{\tau_{n\epsilon}}\right) \nonumber\\
	&\;& + \delta\mu^2\left(\frac{\partial n}{\partial\mu}\frac{1}{\tau_{nn}}+\frac{\partial\epsilon}{\partial\mu}\frac{1}{\tau_{n\epsilon}}\right) + \ldots
\end{eqnarray}
where the dots are the standard dissipative terms. Using the constraints \eqref{eqn:physical_ansatz}, and the results of \eqref{eqn:entropy_constraint_order_one}, we can simplify the above expression \eqref{Eq:Entropy2} to
\begin{align}
	T_{(0)}\partial_\mu \delta s^\mu&= \delta\mu^2\left(\frac{\partial n}{\partial\mu}\frac{1}{\tau_{nn}}+\frac{\partial\epsilon}{\partial\mu}\frac{1}{\tau_{n\epsilon}}\right)+\dots \; . 
\end{align}
The remnant non-zero term can be written in terms of the susceptibilities as
\begin{equation}\label{eqn:constraint_on_taunn}
	\frac{\delta\mu^2}{\tau_{nn}}\left(\chi_{\epsilon\epsilon}\chi_{nn}-\chi_{\epsilon n}^2\right)\lessgtr 0
\end{equation}
where the sign of the inequality depends on the sign of $\frac{\partial\epsilon}{\partial T}$. Because the susceptibility matrix is positive definite, the bracket is also positive. Hence when $\frac{\partial\epsilon}{\partial T}\geq0$ then $\tau_{nn}\geq0$, while if $\frac{\partial\epsilon}{\partial T}<0$ then $\tau_{nn}<0$. Importantly, this condition is not an extra equality-type constraint on $\tau_{nn}$, meaning it is not a fixed parameter in our hydrodynamic model\footnote{Moreover, one could add a second order in fluctuation piece proportional to $\delta \mu^2$ as a charge current or energy relaxation term. Subsequently, we can completely lift any constraints imposed on $\tau_{nn}$ by tuning the coefficient of this term appropriately.}. Notice that for bulk condensed matter systems the specific heat $\frac{\partial\epsilon}{\partial T}$ is generically expected to be positive.}

{\ A final, not necessarily independent, constraint on the relaxation time arises from requiring the linear stability of the modes. In $d+1$ dimensions there are $d+2$ modes, which at zero-wavevector\footnote{The expressions at non-zero wavevector are complicated and dependent on how one chooses to scale the relaxation terms in comparison to the momenta.} are
\begin{equation}
		\label{Eq:LinearisedModes}
		\omega=-\frac{i}{\tau_p} \; , \qquad\omega = - \frac{i}{2} \left( \frac{1}{\tau_{\epsilon \epsilon}} +  \frac{1}{\tau_{nn}} \right) 
				    \pm \frac{i}{2} \sqrt{ \left( \frac{1}{\tau_{\epsilon \epsilon}} -  \frac{1}{\tau_{nn}} \right)^2 + \frac{4}{\tau_{n \epsilon} \tau_{\epsilon n}}} \; ,
\end{equation}
where the first mode has multiplicity $d$. The stability of the state then requires that the imaginary part of these modes be negative. Thus the first mode simply gives us the same constraint, $\tau_p\geq0$, imposed by enforcing the second law. If we liberate ourselves from the second law, but maintain Onsager reciprocity, we have three free parameters and can readily arrange for propagating and/or unstable modes from the second expression \eqref{Eq:LinearisedModes}. On the other hand, employing all our constraints, we find that the second set of modes in \eqref{Eq:LinearisedModes} saturate the linearised stability requirement i.e. 
\begin{equation}
	\omega=0 \; , \qquad\qquad\omega=-i\left(\frac{1}{\tau_{nn}}+\frac{1}{\tau_{\epsilon\epsilon}}\right)=-\frac{i}{\tau_{nn}}\left(\frac{\frac{\partial\epsilon}{\partial T}-\mu_{(0)}\frac{\partial n}{\partial T}}{\frac{\partial\epsilon}{\partial T}}\right) \; . 
\end{equation}
The second expression above gives another constraint on the sign of $\tau_{nn}$ that depends on the thermodynamics, similar to what happens in \eqref{eqn:constraint_on_taunn}.}

{\ While we have determined the one parameter family of relaxations leading to a linearised hydrodynamics respecting Onsager reciprocity and the second law, we should remind ourselves that hydrodynamics is not just a linearised theory. While a full investigation of non-linear corrections is beyond the scope of this paper, it is a reasonable question to ask whether our one-parameter linearised expressions can come from a non-linear formulation. This boils down to a question of integrability in thermodynamics. To investigate it, let us change basis for our relaxations so that
	\begin{equation}
		    \frac{1}{\tau_{\epsilon \epsilon}} \delta \epsilon + \frac{1}{\tau_{\epsilon n}} \delta n		= \frac{1}{\tau_{\epsilon T}} \delta T + \frac{1}{\tau_{\epsilon \mu}} \delta \mu \; , \qquad \frac{1}{\tau_{n \epsilon}} \delta \epsilon + \frac{1}{\tau_{n n}} \delta n		= \frac{1}{\tau_{n T}} \delta T + \frac{1}{\tau_{n \mu}} \delta \mu \; . 
	\end{equation}
Imposing our Onsager reciprocity constraint and the second law on our relaxation terms, we find
	\begin{equation}
		\label{Eq:AlternateRelaxations}
		\frac{1}{\tau_{\epsilon T}} = 0 \; , \qquad \frac{1}{\tau_{n T}} = 0 \; , \qquad \tau_{\epsilon \mu} = \frac{\tau_{nn}}{\mu_{(0)}}\frac{ \frac{\partial \epsilon}{\partial T}}{\frac{\partial n}{\partial \mu} \frac{\partial \epsilon}{\partial T} - \frac{\partial n}{\partial T} \frac{\partial \epsilon}{\partial \mu}} \; , \qquad  \tau_{n \mu} = \frac{\tau_{nn} \frac{\partial \epsilon}{\partial T}}{\frac{\partial n}{\partial \mu} \frac{\partial \epsilon}{\partial T} - \frac{\partial n}{\partial T} \frac{\partial \epsilon}{\partial \mu}} \; . \qquad
	\end{equation}
Suppose there exist a pair of differentiable functions $\Gamma_{\epsilon}$ and $\Gamma_{n}$ at the non-linear level whose linearisations lead to our relaxation terms,
	\begin{subequations}
	\begin{eqnarray}
		\delta \Gamma_{\epsilon} &=& \frac{\partial \Gamma_{\epsilon}}{\partial \mu} \delta \mu +  \frac{\partial \Gamma_{\epsilon}}{\partial \mu} \delta T = \frac{1}{\tau_{\epsilon \mu}} \delta \mu +  \frac{1}{\tau_{\epsilon T}} \delta T  \; , \\
		\delta \Gamma_{n} &=& \frac{\partial \Gamma_{n}}{\partial \mu} \delta \mu +  \frac{\partial \Gamma_{n}}{\partial \mu} \delta T = \frac{1}{\tau_{n\mu}} \delta \mu +  \frac{1}{\tau_{nT}} \delta T  \; .
	\end{eqnarray}
	\end{subequations}
It follows that our linearised relaxations are required to satisfy the integrability conditions,
	\begin{equation}
		\label{Eq:Integrability}
		 	 \frac{\partial}{\partial T} \left( \frac{1}{\tau_{\epsilon \mu}} \right)
		 = \frac{\partial}{\partial \mu} 	\left( \frac{1}{\tau_{\epsilon T}} \right) \; , \qquad
		 	\frac{\partial}{\partial T} \left( \frac{1}{\tau_{n \mu}} \right)
		 = \frac{\partial}{\partial \mu} 	\left( \frac{1}{\tau_{n T}} \right) \; ,
	\end{equation}
which come from commutivity of second derivatives on $\Gamma_{\epsilon}$ and $\Gamma_{n}$\footnote{It is clear that no equivalent $\Gamma_{p}$ can exist.}. From \eqref{Eq:AlternateRelaxations} we see that
	\begin{eqnarray}
		 \frac{\partial}{\partial \mu} 	\left( \frac{1}{\tau_{\epsilon T}} \right) = 0 \; , \qquad \frac{\partial}{\partial \mu} 	\left( \frac{1}{\tau_{n T}} \right) =0 \; , 
	\end{eqnarray}
and plugging this result into \eqref{Eq:Integrability}, we find $\tau_{\epsilon \mu}$ and $\tau_{n \mu}$ are independent of $T$. Therefore, the most general $\tau_{nn}$ compatible with the existence of $\Gamma_{\epsilon}$ and $\Gamma_{n}$ are
	\begin{equation}
		\frac{1}{\tau_{nn}(T,\mu)} = \frac{f(\mu) \frac{\partial \epsilon}{\partial T}}{\frac{\partial n}{\partial \mu} \frac{\partial \epsilon}{\partial T} - \frac{\partial n}{\partial T} \frac{\partial \epsilon}{\partial \mu}}
	\end{equation}
for $f$ an arbitrary function of $\mu$, while all other relaxations are fixed by our constraints. We repeat that this final result is a consequence of Onsager reciprocity, positivity of entropy production and the existence of a non-linear lift of the linearised quasi-hydrodynamic model. Sacrificing even one of the requirements leads to a much more general result.}

{To conclude this section, we comment that the fluid with relaxations described above is quite different from the one presented in \cite{Amoretti:non_dissipative}. There we considered a complete (non-linear) theory of hydrodynamics, in the hydrostatic regime, while the relaxations considered here are defined to be out-of-equilibrium quantities. Moreover the constraint obtained in \cite{Amoretti:non_dissipative} between energy $\Gamma_\epsilon$ and momentum relaxation $\Gamma_p$, i.e.
	\begin{equation}
		\Gamma_\epsilon=\Gamma_pp^iv_i \; , 
	\end{equation}
	ensures that at linear order in small velocity, energy is still conserved and only momentum decays with the usual constraint $\tau_p^{-1}=\Gamma_p\geq0$.}

\section{The background field method and time-reversal covariance}\label{sec:variational_approach}

{\noindent In the previous section, we saw how imposing Onsager reciprocity on the correlators between conserved charges, constrains the possible form of charge relaxation for an example charged relativistic fluid. In the present section, we explore whether additional constraints appear when we extend time-reversal covariance to the correlators between the full current and energy-momentum tensor. 

To evaluate the complete correlators, we use the background field approach. In the background field approach, one places the fluid on a curved background $g_{\mu\nu}$ with a generic gauge field $A_\mu$ and defines the expectation values
	\begin{equation}
		\mathcal{J}^{\mu}(x) = \sqrt{-g} \langle J^{\mu}(x) \rangle_{A,g} \; , \qquad \mathcal{T}^{\mu \nu}(x) = \sqrt{-g} \langle T^{\mu \nu}(x) \rangle_{A,g}~.
	\end{equation}
On the RHS of each equality, the expectation values of $J^\mu$ and $T^{\mu\nu}$ in the presence of $A_\mu$ and $g_{\mu\nu}$ are given by the on-shell values of $J^\mu$ and $T^{\mu\nu}$. This point of view allows us to define the retarded correlators of the stress tensor and current via varying with respect to the sources $A_\mu$ and $g_{\mu\nu}$ and then taking the flat space limit \cite{Kovtun:lectures}, i.e.
	\begin{subequations}
	\label{Eq:VariationalCorr}
	\begin{align}
	& \langle J^{\mu} J^{\nu} \rangle_{\mathrm{R}}(x) = - \left. \frac{\delta \mathcal{J}^{\mu}(x)}{\delta A_{\nu}(0)} \right|_{g=\eta,A=0} \; , \qquad & \langle T^{\mu \nu} J^{\rho} \rangle_{\mathrm{R}}(x) = - \left. \frac{\delta \mathcal{T}^{\mu \nu}(x)}{\delta A_{\rho}(0)} \right|_{g=\eta,A=0} \; , \qquad \\
		& \langle J^{\mu} T^{\nu \rho} \rangle_{\mathrm{R}}(x) = -  \left. 2 \frac{\delta \mathcal{J}^{\mu}(x)}{\delta g_{\nu \rho}(0)} \right|_{g=\eta,A=0} \; , \qquad & \langle T^{\mu \nu} T^{\rho \sigma} \rangle_{\mathrm{R}}(x) = - \left. 2 \frac{\delta \mathcal{T}^{\mu \nu}(x)}{\delta g_{\rho \sigma}(0)} \right|_{g=\eta,A=0} \; .  \qquad
	\end{align}
	\end{subequations}  
	We see that by solving the linearised hydrodynamic equations for the evolution of the hydrodynamic fields in terms of the external sources, this method gives direct access to all the correlators of the stress tensor and charge current.} 

{\ We consider now the same theory of linearised hydrodynamics discussed in the previous section, but this time placed on a curved background. The non-linear Landau-frame constitutive relations are, up to order one in derivatives,
\begin{subequations}
	\label{Eq:Nonlinearhydro}
	\begin{align}
		T^{\mu \nu} &= \epsilon u^\mu u^\nu+P \Delta^{\mu\nu}-\eta\Delta^{\mu\alpha}\Delta^{\nu\beta}\left(\nabla_\alpha u_\beta+\nabla_\beta u_\alpha-\frac{2}{d}g_{\alpha\beta}\nabla_\lambda u^\lambda\right)-\zeta\nabla_\mu u^\mu+\mathcal{O}(\nabla^2) \; , \\
		J^\mu &= nu^\mu+\sigma\Delta^{\mu\nu}\left(E_\nu-T\nabla_\nu\frac{\mu}{T}\right)+\mathcal{O}(\nabla^2) \; , 
	\end{align}
\end{subequations}
where $\Delta^{\mu\nu}=g^{\mu\nu}+u^\mu u^\nu$ is the projector transverse to the velocity and $E_\mu=F_{\mu\nu}u^\nu$ the electric field. As we are looking at linearised hydrodynamics around global thermodynamic equilibrium, we introduce a preferred time-like direction $\tau_{\mu}(x)$ such that $\tau^{\mu} \tau_{\mu} = -1$. 
 Notice that while $\tau_{\mu}(x)$ is coordinate covariant, in Cartesian coordinates in $d+1$ dimensional Minkowski space we can choose for it to be $\tau_{\mu} = (-1,\vect{0})$. Consequently, $\nabla_{\mu}^{(0)} \tau^{\mu}=0$ in all coordinate systems parameterising Minkowski space.}

{\ We can now linearise the theory around an equilibrium configuration with constant temperature, constant chemical potential, zero fluid velocity and flat spacetime
\begin{subequations}
	\begin{align}
		T &= T_{(0)}+\delta T \; , 	&	\mu &= \mu_{(0)}+\delta\mu\; , & u^\mu &=\tau^\mu+\delta v^\mu \; ,	\\
		g^{\mu\nu}&=\eta^{\mu\nu}+\delta h^{\mu\nu} \; , &    A^\mu&=\delta A^\mu \; . 
	\end{align}
\end{subequations}
The (linearised) equations of motion read 
\begin{subequations}\label{eqn:covariant_equations_of_motion}
	\begin{align}
		\label{Eq:EnergyNonCon}
		&\tau^\mu\partial_\mu\delta\epsilon + \left( \epsilon_{(0)} + P_{(0)} \right) \nabla^{(0)}_{\mu} \delta v^{\mu} = - \left( \frac{1}{\tau_{\epsilon \epsilon}} \delta \epsilon +  \frac{1}{\tau_{\epsilon n}} \delta n \right)  \; , \\
		\label{Eq:ChargeNonCon}
		&\tau^\mu\partial_\mu\delta n + n_{(0)} \nabla^{(0)}_{\mu} \delta v^{\mu} - \sigma \left( P^{\mu \nu} \nabla^{(0)}_{\mu} \partial_{\nu} \delta \mu - \frac{\mu_{(0)}}{T_{(0)}} P^{\mu \nu} \nabla^{(0)}_{\mu} \partial_{\nu} \delta T -\nabla^{(0)}_{\mu} \delta E^{\mu}\right) \nonumber \\
		&= - \left( \frac{1}{\tau_{n \epsilon}} \delta \epsilon +  \frac{1}{\tau_{nn}} \delta n \right)\; , \\
		\label{Eq:MomentumNonCon}
		&\; P^{\mu \nu} \partial_{\nu} \delta P
		-  \zeta P^{\mu \nu} \nabla_{\nu}^{(0)} \nabla_{\rho}^{(0)} \delta v^{\rho} + \left( \epsilon_{(0)} + P_{(0)} \right) P\indices{^{\mu}_{\rho}} \left( \tau^{\nu} \nabla_{\nu}^{(0)} \delta v^{\rho}  \right) - 2 \eta P^{\mu \rho} \nabla_{\sigma}^{(0)} \delta \sigma\indices{^\sigma_\rho} \nonumber \\
		&= - \frac{1}{\tau_{p}} \left( \epsilon_{(0)} + P_{(0)} \right) \delta v^{\mu} + n_{(0)} \delta E^{\mu} - P_{(0)} P\indices{^{\mu}_{\nu}} \nabla_{\rho}^{(0)} \delta h^{\rho \nu} - \left( \epsilon_{(0)} + P_{(0)} \right) P\indices{^{\mu}_{\rho}}  \tau^{\nu} \delta \Gamma_{\nu \sigma}^{\rho} \tau^{\sigma}~,
	\end{align}
\end{subequations}
where $P^{\mu\nu}=\eta^{\mu\nu}+\tau^\mu\tau^\nu$ is the projector transverse to the background velocity.  As anticipated in the introduction, the relaxation terms explicitly break the Lorentz symmetries of the theory. For this reason we have been agnostic on possible ``covariantizations'' of these terms in writing the equations of motion, simply writing the relaxation terms as they appear in flat spacetime. We emphasise that our expressions are explicitly coordinate invariant by use of $\nabla^{(0)}_\mu$ instead of $\partial_\mu$ for the background covariant derivative which accounts for the existence of non-linearised connections on the flat background if one uses a curvilinear coordinate system.}

{\ Given the above setup, it is not difficult to show that there are correlators which do not satisfy the time-reversal covariance condition \eqref{eqn:Onsager_relations}. For example,
	\begin{multline}
		\left. \langle T^{tt} T^{xx} \rangle - \langle T^{xx} T^{tt} \rangle \right|_{\vect{k}=0} = \\
		=-\frac{\left(\left(\epsilon_{(0)}+P_{(0)}\right)\left(\tau_{n\epsilon}\tau_{\epsilon n}-\tau_{nn}\tau_{\epsilon\epsilon}\right)\right)+i\tau_{nn}\tau_{n\epsilon}\left((\epsilon_{(0)}+P_{(0)})\tau_{\epsilon n}+n_{(0)}\tau_{\epsilon\epsilon}\right)\omega}{\tau_{nn}\tau_{\epsilon\epsilon}+\tau_{n\epsilon}\tau_{\epsilon n}\left(i+\tau_{nn}\omega\right)\left(i+\tau_{\epsilon\epsilon}\omega\right)}
	\end{multline}
This remains the case, even when we identify the energy and charge relaxation terms with those preserving the Onsager relations (and the second law) in the Martin-Kadanoff approach, \eqref{eqn:Onsager_general}. In addition, the correlators are generically different from the ones obtained in the previous section.\footnote{Even after taking into account the fact that the two approaches may differ by contact terms.}}

{\ If we want the correlators to be time-reversal covariant, then we must modify some aspect of our hydrodynamic formulation. We choose to modify the hydrodynamic equations, by including additional source terms in the hydrodynamic equations that vanish when the metric and gauge field take their background values. That is, we include additional sources to the hydrodynamic equations depending  non-trivially on the differences $\delta(g_{\mu\nu} - \eta_{\mu\nu})$ and $\delta(A_\mu -A^{(0)}_\mu)$. The relaxations can then be understood as explicitly breaking the background independence of the theory, which manifests in a preferred metric and gauge field. This is not to say that one cannot place this relaxed hydrodynamics on a curved background, only that the equation of motion depends explicitly on the background metric. It is also important to reiterate that the resultant theory is coordinate invariant, even if it is not background independent.}

{\ To employ our method, we proceed by brute force and compute the correlators with arbitrary additional source terms constructed from $\delta h_{\mu \nu}$ and $\delta A_{\mu}$. In particular, we schematically write\footnote{We have also confirmed if one includes relaxation terms that have explicit derivatives of the hydrodynamic variables one is also required to add derivatives of $\delta h_{\mu \nu}$ and $\delta A_{\mu}$ to the linearised equation of motion. }
\begin{equation}\label{eqn:extra_source_terms}
		\left( \mathrm{sources \; of \; \eqref{eqn:covariant_equations_of_motion}} \right)  \rightarrow  \left( \mathrm{sources \; of \; \eqref{eqn:covariant_equations_of_motion}} \right) + c\indices{_{\alpha}^{\mu \nu}} \delta h_{\mu \nu} + r\indices{_{\alpha}^{\mu}} \delta A_{\mu} 
\end{equation}
and compute the two point functions using these modified equations of motion via the variational approach. In the above expression $\alpha=\{\epsilon,n,x,y,z\}$ identifies the relevant equation of motion. Imposing time-reversal covariance on these Green's functions gives us a set of relations that must be solved for the arbitrary coefficients in \eqref{eqn:extra_source_terms}. We do this for the full correlator obtained from the background field method, rather than the truncated correlator (where one accounts for and excises any spurious higher derivative terms). Consequently our complete correlators respect time-reversal covariance.}

{\ There are in total $70$ source terms that we can add, however, since there are no parity-breaking operators in the theory, $c\indices{_{\alpha}^{\mu \nu}}, r\indices{_{\alpha}^{\mu}}$ are parity even and this allows us to use rotational invariance and parity with respect to a single axis (which itself is a consequence of parity and rotational invariance) $\mathcal{P}_i:i\rightarrow-i$ ($i=x,y,z$) to reduce the number of sources to only 13. Because $\tau_p$ has the same value in all the directions, we also make that ansatz the sources are isotropic, which reduces their number down to $9$. Finally, after imposing time-reversal covariance\footnote{On a practical level: first we imposed time-reversal covariance relations at $\vect{k}=0$, then at $\omega=k_y=k_z=0$. These two sets of constraints are enough to identify all the coefficients. We subsequently check explicitly that the full correlators satisfy time-reversal covariance.} on the full correlators we end up with only 4 new source terms, while all the other coefficients are found to be zero. The only equations of motion that receive corrections are those for energy and charge, i.e.
\begin{subequations}\label{eqn:final_relaxations}
	\begin{align}
		\text{energy:}&	&	&- \left( \frac{1}{\tau_{\epsilon \epsilon}} \delta \epsilon +  \frac{1}{\tau_{\epsilon n}} \delta n \right)-c\indices{_{\epsilon}^{tt}}\delta h_{tt}-r\indices{_{\epsilon}^{t}}\delta A_t~,
		\\
		\text{charge:}&	&	&- \left( \frac{1}{\tau_{n \epsilon}} \delta \epsilon +  \frac{1}{\tau_{nn}} \delta n \right)-c\indices{_{n}^{tt}}\delta h_{tt}-r\indices{_{n}^{t}}\delta A_t~,
	\end{align}
\end{subequations}
where the value of the coefficients is fixed in terms of the standard relaxations
\begin{subequations}
	\label{Eq:CorrectedConservation}
	\begin{align}
		c\indices{_{\epsilon}^{tt}}&=\frac{1}{2}\left(\frac{\chi_{\epsilon n}}{\tau_{\epsilon n}}+\frac{\chi_{\epsilon\epsilon}}{\tau_{\epsilon\epsilon}}\right)~,
		\\
		r\indices{_{\epsilon}^{t}}&=\frac{\chi_{\epsilon\epsilon}}{\tau_{n\epsilon}}+\frac{\chi_{\epsilon n}}{\tau_{nn}}=2c\indices{_{n}^{tt}}~,
		\\
		c\indices{_{n}^{tt}}&=\frac{1}{2}\left(\frac{\chi_{\epsilon\epsilon}}{\tau_{n\epsilon}}+\frac{\chi_{\epsilon n}}{\tau_{nn}}\right)~,
		\\
		r\indices{_{n}^{t}}&=\frac{\chi_{\epsilon n}}{\tau_{n\epsilon}}+\frac{\chi_{nn}}{\tau_{nn}}~.
	\end{align}
\end{subequations}
There are no additional constraints on the relaxation time parameters $\tau$ beyond those imposed in the Martin-Kadanoff procedure \eqref{eqn:Onsager_general}. Hence we have confirmed that i) relaxed hydrodynamics can be made time-reversal covariant and ii) that the correlators obtained by the variational procedure including the $c$s and $r$s agree with those obtained by the Martin-Kadanoff one, up to the usual contact terms\footnote{We can also compare our results to those of detailed in the appendix ``Coupling to external sources'' of \cite{Delacretaz:2021qqu} (see also \cite{Armas:approximate_symmetries}). The authors compute the constitutive relations from the Schwinger-Keldysh formalism of relaxed hydrodynamics without a stress tensor but with a pseudo-Goldstone field (which we can set to zero without issue). They find the presence of an additional term proportional to the time-component of the gauge field as we have added. This is unsurprising as the Schwinger-Keldysh formalism has time-reversal covariance and positivity of entropy production built in.}. As the same exact procedure presented above can be applied to a perfect fluid, leading to the same results \eqref{Eq:CorrectedConservation}, our expressions are hydrodynamic frame covariant as one would expect of a physically meaningful theory.}

As we mentioned above, only energy and charge equations receive corrections, however this is due to our specific choice of how to express the relaxations. Picking different ``covariantizations'' to write the relaxations (e.g. $\delta n=\tau_\mu\delta J^\mu$) in \eqref{eqn:covariant_equations_of_motion}, will in general give different values for the coefficients $r$s and $c$s, but will be such that the final equations of motion are unmodified \eqref{eqn:final_relaxations}.

{\ Finally, one can reconsider positivity of entropy production at linear order in fluctuations in light of our new metric and gauge field fluctuation terms. On a weakly curved background the equivalent expression to \eqref{Eq:LowestEntropy} is
	\begin{align}
		T_{(0)}\nabla_\mu^{(0)} \delta s^\mu&=\delta\epsilon\left(\frac{\mu_{(0)}}{\tau_{n\epsilon}}-\frac{1}{\tau_{\epsilon\epsilon}}\right)+\delta n\left(\frac{\mu_{(0)}}{\tau_{nn}}-\frac{1}{\tau_{\epsilon n}}\right)\\
		\;&\quad - \left( r\indices{_{\epsilon}^{t}} - \mu_{(0)} r\indices{_{n}^{t}} \right) \tau^{\mu} \delta A_{\mu} - \left( c\indices{_{\epsilon}^{tt}} - \mu_{(0)} c\indices{_{n}^{tt}} \right) \tau^{\mu} \delta g_{\mu \nu} \tau^{\nu}\nonumber \\
		\;&\quad + \mathcal{O}(\delta^2,\partial^2)
	\end{align}
As we have not added second order in fluctuation additional terms, this is the only expression we need to consider. Somewhat miraculously, imposing the entropy positivity conditions we found in section \ref{sec:Martin-Kadanoff}, \eqref{eqn:entropy_constraint_order_one} and \eqref{eqn:physical_ansatz}, also happens to set the new terms to zero. Thus, our inclusion of background-dependent terms  to the energy and charge conservation equations does not violate the second law to order one in fluctuations.}

\section{Discussion}

{\noindent In this paper we have analyzed the implications of time-reversal covariance of the microscopic theory, i.e. Onsager-Casimir relations, on a theory of linearised relativistic hydrodynamic in the presence of generic relaxations. We found a set of constraints that the relaxation parameters must obey in order for the fluid to satisfy Onsager relations \eqref{eqn:Onsager_general}, positivity of entropy production \eqref{eqn:physical_ansatz} and linearised stability \eqref{Eq:LinearisedModes}, which reduce the number of free relaxation parameters to only one.}

{\ Subsequently we have computed all the retarded two-point functions for such a fluid by considering small perturbations of the metric and gauge field: we found that in general this method gives Green functions that are not time-reversal covariant and do not match with the ones obtained by linear response theory. One of the core results of this work is to show that it is possible to overcome these problems by considering extra source terms in the equations of motion and, surprisingly, the coefficients of these source terms are completely fixed by the simple requirement of time-reversal covariance, leading to the final result \eqref{Eq:CorrectedConservation}.}

{\ Although we tested this procedure only on a relativistic fluid with a $U(1)$ symmetry, which is the simplest one to couple to curved spacetime, we expect the same method to also work for other hydrodynamic theories with different spacetime and internal symmetries. All the more so as our constraints can be derived at the ideal level. Furthermore, this approach should work also for non-trivial equilibrium backgrounds, e.g. a constant magnetic field, a curved spacetime or in the presence of topological terms such as Chern-Simon's terms \cite{Brattan:2013wya,Brattan:2014moa}. To check the validity of this claim could be the goal of succeeding works and it would be quite interesting to find a situation where time-reversal covariance and positivity of entropy production are not sufficient to fix the extra variational terms.}

{\ Regarding future perspectives, it would be interesting to study how these relaxations can be consistently included in the quasi-hydrodynamic description beyond the linearised regime. In particular, the presence of relaxation terms in the equations of motion could induce new transport coefficients in the constitutive relations or modify the values of known ones. One could also consider hydrodynamic $N$-point functions and ascertain whether time reversal covariance is sufficient to fix higher order in fluctuation terms in the effective hydrodynamic equations.}

{\ It is also possible, with the findings of this work, to re-analyze certain known results related to the transport properties of anomalous hydrodynamics, e.g. in the context of studying the thermoelectric properties of Weyl semimetals. Specifically, as already mentioned in the introduction, generic relaxations are needed to obtain finite DC conductivities for a anomalous fluids \cite{Landsteiner:negative_magnetoresistivity,Abbasi:magneto-transport,Lucas:hydrodynamic_theory}, hence it would be fruitful to apply the methods developed here to study the DC limit of the conductivities presented in \cite{Amoretti:frame_dependence}.}

{\ Finally, it could be interesting to study models of relaxed hydrodynamics in the context of kinetic theory \cite{Fritz:hydrodynamic_electronic_transport}, holography \cite{Landsteiner:anomalous_magnetoconductivity_holography} or using the Schwinger-Keldysh EFT formalism \cite{Glorioso:lectures}. Because these approaches often impose different constraints on the EFT compared to this work\footnote{For example, the Schwinger-Keldysh formalism requires the dynamical KMS condition and unitarity, which in turn automatically impose non-linear Onsager relations and positivity of entropy production}, they could give insight in order to check the universality of our results and how they are realised in different contexts.}

\section*{Acknowledgements}

{\noindent We would like to acknowledge helpful discussions with Pavel Kovtun, Francisco Pe{\~n}a-Benitez, Amos Yarom and Vaios Ziogas. A.A. and I.M. have been partially supported by the “Curiosity Driven Grant 2020” of the University of Genoa and the INFN Scientific Initiative SFT: “Statistical Field Theory, Low-Dimensional Systems, Integrable Models and Applications”. This project has also received funding from the European Union’s Horizon 2020 research and innovation programme under the Marie Sk\l{}odowska-Curie grant agreement No. 101030915. }

\bibliography{refs}

\providecommand{\href}[2]{#2}\begingroup\raggedright\begin{thebibliography}{10}

\bibitem{Hartnoll:theory}
S.~A. Hartnoll, P.~K. Kovtun, M.~Muller, and S.~Sachdev, {\it {Theory of the
  Nernst effect near quantum phase transitions in condensed matter, and in
  dyonic black holes}},  {\em Phys. Rev. B} {\bf 76} (2007) 144502,
  [\href{http://arxiv.org/abs/0706.3215}{{\tt arXiv:0706.3215}}].

\bibitem{Andrade:simple_holographic_model}
T.~Andrade and B.~Withers, {\it {A simple holographic model of momentum
  relaxation}},  {\em JHEP} {\bf 05} (2014) 101,
  [\href{http://arxiv.org/abs/1311.5157}{{\tt arXiv:1311.5157}}].

\bibitem{Amoretti:2017xto}
A.~Amoretti, A.~Braggio, N.~Maggiore, and N.~Magnoli, {\it {Thermo-electric
  transport in gauge/gravity models}},  {\em Adv. Phys. X} {\bf 2} (2017),
  no.~2 409--427.

\bibitem{Amoretti:2014zha}
A.~Amoretti, A.~Braggio, N.~Maggiore, N.~Magnoli, and D.~Musso, {\it
  {Thermo-electric transport in gauge/gravity models with momentum
  dissipation}},  {\em JHEP} {\bf 09} (2014) 160,
  [\href{http://arxiv.org/abs/1406.4134}{{\tt arXiv:1406.4134}}].

\bibitem{Amoretti:magneto-transport_holography}
A.~Amoretti and D.~Musso, {\it {Magneto-transport from momentum dissipating
  holography}},  {\em JHEP} {\bf 09} (2015) 094,
  [\href{http://arxiv.org/abs/1502.02631}{{\tt arXiv:1502.02631}}].

\bibitem{Landsteiner:negative_magnetoresistivity}
K.~Landsteiner, Y.~Liu, and Y.-W. Sun, {\it {Negative magnetoresistivity in
  chiral fluids and holography}},  {\em JHEP} {\bf 03} (2015) 127,
  [\href{http://arxiv.org/abs/1410.6399}{{\tt arXiv:1410.6399}}].

\bibitem{Abbasi:magneto-transport}
N.~Abbasi, A.~Ghazi, F.~Taghinavaz, and O.~Tavakol, {\it {Magneto-transport in
  an anomalous fluid with weakly broken symmetries, in weak and strong
  regime}},  {\em JHEP} {\bf 05} (2019) 206,
  [\href{http://arxiv.org/abs/1812.11310}{{\tt arXiv:1812.11310}}].

\bibitem{Lucas:hydrodynamic_theory}
A.~Lucas, R.~A. Davison, and S.~Sachdev, {\it {Hydrodynamic theory of
  thermoelectric transport and negative magnetoresistance in Weyl semimetals}},
   {\em Proc. Nat. Acad. Sci.} {\bf 113} (2016) 9463,
  [\href{http://arxiv.org/abs/1604.08598}{{\tt arXiv:1604.08598}}].

\bibitem{Landsteiner:anomalous_magnetoconductivity_holography}
A.~Jimenez-Alba, K.~Landsteiner, Y.~Liu, and Y.-W. Sun, {\it {Anomalous
  magnetoconductivity and relaxation times in holography}},  {\em JHEP} {\bf
  07} (2015) 117, [\href{http://arxiv.org/abs/1504.06566}{{\tt
  arXiv:1504.06566}}].

\bibitem{Pongsangangan:hydrodynamics_charged}
K.~Pongsangangan, T.~Ludwig, H.~T.~C. Stoof, and L.~Fritz, {\it Hydrodynamics
  of charged two-dimensional dirac systems. i. thermoelectric transport},  {\em
  Phys. Rev. B} {\bf 106} (Nov, 2022) 205126,
  [\href{http://arxiv.org/abs/2206.09687}{{\tt arXiv:2206.09687}}].

\bibitem{Fritz:hydrodynamic_electronic_transport}
L.~Fritz and T.~Scaffidi, {\it {Hydrodynamic electronic transport}},
  \href{http://arxiv.org/abs/2303.14205}{{\tt arXiv:2303.14205}}.

\bibitem{Narozhny:hydrodynamic_approach}
B.~N. Narozhny and I.~V. Gornyi, {\it Hydrodynamic approach to electronic
  transport in graphene: Energy relaxation},  {\em Frontiers in Physics} {\bf
  9} (2021) [\href{http://arxiv.org/abs/2102.00207}{{\tt arXiv:2102.00207}}].

\bibitem{Gall:electronic_viscosity}
V.~Gall, B.~N. Narozhny, and I.~V. Gornyi, {\it Electronic viscosity and energy
  relaxation in neutral graphene},  {\em Phys. Rev. B} {\bf 107} (Jan, 2023)
  045413, [\href{http://arxiv.org/abs/2206.07414}{{\tt arXiv:2206.07414}}].

\bibitem{Amoretti:non_dissipative}
A.~Amoretti, D.~K. Brattan, L.~Martinoia, and I.~Matthaiakakis, {\it
  {Non-dissipative electrically driven fluids}},
  \href{http://arxiv.org/abs/2211.05791}{{\tt arXiv:2211.05791}}.

\bibitem{Kovtun:lectures}
P.~Kovtun, {\it {Lectures on hydrodynamic fluctuations in relativistic
  theories}},  {\em J. Phys. A} {\bf 45} (2012) 473001,
  [\href{http://arxiv.org/abs/1205.5040}{{\tt arXiv:1205.5040}}].

\bibitem{Kadanoff:hydrodynamics}
L.~P. Kadanoff and P.~C. Martin, {\it Hydrodynamic equations and correlation
  functions},  {\em Annals of Physics} {\bf 24} (1963) 419--469.

\bibitem{Note1}
Notice that it is not possible to simply swap each $\omega $ with $\omega
  +\protect \frac {i}{\tau }$ in the Green's functions.

\bibitem{Note2}
This is even more relevant in the presence of energy and charge relaxation,
  because in this scenario we cannot naively use Ward identities to relate e.g.
  $\langle J^t J^t\rangle $ to $\langle J^i J^i\rangle $.

\bibitem{Note3}
Recall the susceptibility matrix is symmetric, $\chi _{ab} = \chi _{ba}$.

\bibitem{Note4}
Our relaxations will break Lorentz invariance and one may in principle expect
  new transport coefficients to appear depending on how the microscopic theory
  couples to processes responsible for breaking this symmetry. These new
  coefficients will not change results related to our relaxations and we can,
  without loss of generality, assume that they happen to be zero for our fluid
  henceforth.

\bibitem{Note5}
We could also express the second law in terms of $\delta T$ and $\delta \mu $.
  This leads to a $2\times 2$ linear system of equations that has a solution
  with ${\protect \rm det}\chi = 0$. In what follows, we ignore this solution
  as unphysical.

\bibitem{Amoretti:hydrodynamic_magneto-transport}
A.~Amoretti, D.~Arean, D.~K. Brattan, and N.~Magnoli, {\it {Hydrodynamic
  magneto-transport in charge density wave states}},  {\em JHEP} {\bf 05}
  (2021) 027, [\href{http://arxiv.org/abs/2101.05343}{{\tt arXiv:2101.05343}}].

\bibitem{Amoretti:2021lll}
A.~Amoretti, D.~Arean, D.~K. Brattan, and L.~Martinoia, {\it {Hydrodynamic
  magneto-transport in holographic charge density wave states}},  {\em JHEP}
  {\bf 11} (2021) 011, [\href{http://arxiv.org/abs/2107.00519}{{\tt
  arXiv:2107.00519}}].

\bibitem{Amoretti:2022acb}
A.~Amoretti and D.~K. Brattan, {\it {On the hydrodynamics of (2 +
  1)-dimensional strongly coupled relativistic theories in an external magnetic
  field}},  {\em Mod. Phys. Lett. A} {\bf 37} (2022), no.~21 2230010,
  [\href{http://arxiv.org/abs/2209.11589}{{\tt arXiv:2209.11589}}].

\bibitem{Note6}
Moreover, one could add a second order in fluctuation piece proportional to
  $\delta \mu ^2$ as a charge current or energy relaxation term. Subsequently,
  we can completely lift any constraints imposed on $\tau _{nn}$ by tuning the
  coefficient of this term appropriately.

\bibitem{Note7}
The expressions at non-zero wavevector are complicated and dependent on how one
  chooses to scale the relaxation terms in comparison to the momenta.

\bibitem{Note8}
It is clear that no equivalent $\Gamma _{p}$ can exist.

\bibitem{Note9}
Even after taking into account the fact that the two approaches may differ by
  contact terms.

\bibitem{Note10}
We have also confirmed if one includes relaxation terms that have explicit
  derivatives of the hydrodynamic variables one is also required to add
  derivatives of $\delta h_{\mu \nu }$ and $\delta A_{\mu }$ to the linearised
  equation of motion.

\bibitem{Note11}
On a practical level: first we imposed time-reversal covariance relations at
  $\protect \boldsymbol {\protect \mathbf {k}}=0$, then at $\omega =k_y=k_z=0$.
  These two sets of constraints are enough to identify all the coefficients. We
  subsequently check explicitly that the full correlators satisfy time-reversal
  covariance.

\bibitem{Note12}
We can also compare our results to those of detailed in the appendix ``Coupling
  to external sources'' of \cite {Delacretaz:2021qqu} (see also \cite
  {Armas:approximate_symmetries}). The authors compute the constitutive
  relations from the Schwinger-Keldysh formalism of relaxed hydrodynamics
  without a stress tensor but with a pseudo-Goldstone field (which we can set
  to zero without issue). They find the presence of an additional term
  proportional to the time-component of the gauge field as we have added. This
  is unsurprising as the Schwinger-Keldysh formalism has time-reversal
  covariance and positivity of entropy production built in.

\bibitem{Brattan:2013wya}
D.~K. Brattan and G.~Lifschytz, {\it {Holographic plasma and anyonic fluids}},
  {\em JHEP} {\bf 02} (2014) 090, [\href{http://arxiv.org/abs/1310.2610}{{\tt
  arXiv:1310.2610}}].

\bibitem{Brattan:2014moa}
D.~K. Brattan, {\it {A strongly coupled anyon material}},  {\em JHEP} {\bf 11}
  (2015) 214, [\href{http://arxiv.org/abs/1412.1489}{{\tt arXiv:1412.1489}}].

\bibitem{Amoretti:frame_dependence}
A.~Amoretti, D.~K. Brattan, L.~Martinoia, and I.~Matthaiakakis, {\it {On the
  frame dependence of conductivities in anomalous hydrodynamics}},
  \href{http://arxiv.org/abs/2212.09761}{{\tt arXiv:2212.09761}}.

\bibitem{Glorioso:lectures}
P.~Glorioso and H.~Liu, {\it Lectures on non-equilibrium effective field
  theories and fluctuating hydrodynamics},
  \href{http://arxiv.org/abs/1805.09331}{{\tt arXiv:1805.09331}}.

\bibitem{Note13}
For example, the Schwinger-Keldysh formalism requires the dynamical KMS
  condition and unitarity, which in turn automatically impose non-linear
  Onsager relations and positivity of entropy production.

\bibitem{Delacretaz:2021qqu}
L.~V. Delacr\'etaz, B.~Gout\'eraux, and V.~Ziogas, {\it {Damping of
  Pseudo-Goldstone Fields}},  {\em Phys. Rev. Lett.} {\bf 128} (2022), no.~14
  141601, [\href{http://arxiv.org/abs/2111.13459}{{\tt arXiv:2111.13459}}].

\bibitem{Armas:approximate_symmetries}
J.~Armas, A.~Jain, and R.~Lier, {\it Approximate symmetries, pseudo-goldstones,
  and the second law of thermodynamics},
  \href{http://arxiv.org/abs/2112.14373}{{\tt arXiv:2112.14373}}.

\end{thebibliography}\endgroup
\bibliographystyle{JHEP}

\end{document}